\UseRawInputEncoding
\documentclass[prr,twocolumn,superscriptaddress,floatfix,nopacs,longbibliography]{revtex4-2}
\usepackage{graphicx,amsfonts,amssymb,amsmath,hyperref,enumerate,footnote}
\usepackage{graphicx}
\usepackage{float}
\usepackage{indentfirst}
\usepackage{verbatim}
\usepackage{tikz}
\usepackage{ragged2e}
\usetikzlibrary{quantikz}
\usetikzlibrary{positioning}
\usepackage{array}
\usepackage[utf8]{inputenc}
\usepackage{mathtools}
\usepackage{subcaption}
\usepackage{blindtext}
\usepackage{xcolor}
\usepackage{tikz}
\usetikzlibrary{arrows.meta,
                chains,
                positioning,
                shapes.geometric}
\usepackage{siunitx}
\tikzstyle{startstop} = [rectangle, rounded corners, minimum width=3cm, minimum height=1cm,text centered, draw=black, fill=red!30]
\tikzstyle{io} = [trapezium, trapezium left angle=70, trapezium right angle=110, minimum width=3cm, minimum height=1cm, text centered, draw=black, fill=blue!30]
\tikzstyle{process} = [rectangle, minimum width=3cm, minimum height=1cm, text centered, draw=black, fill=orange!30]
\tikzstyle{decision} = [diamond, minimum width=3cm, minimum height=1cm, text centered, draw=black, fill=green!30]
\tikzstyle{arrow} = [thick,->,>=stealth]

\newif\ifhyper
\hypertrue
\ifhyper
\hypersetup{
   citecolor = {green},
   colorlinks = {true}, 
   urlcolor = {blue} 
}
\fi

\newcommand{\beq}{\begin{equation}}
\newcommand{\eeq}{\end{equation}}
\newcommand{\beqa}{\begin{eqnarray}}
\newcommand{\eeqa}{\end{eqnarray}}

\newcommand{\rom}[1]{{{\color{black}{#1}}}}

\def\ket#1{\vert#1\rangle}

\def\Longarrow{\protect\@lra}
\def\@lra{\relbar\joinrel\relbar\joinrel\relbar\joinrel%
          \relbar\joinrel\rightarrow}

\begin{document}

\title{Efficient tensor network simulation of IBM's largest quantum processors}

\author{Siddhartha Patra}
\affiliation{Donostia International Physics Center, Paseo Manuel de Lardizabal 4, E-20018 San Sebasti\'an, Spain}

\author{Saeed S. Jahromi}
\affiliation{Donostia International Physics Center, Paseo Manuel de Lardizabal 4, E-20018 San Sebasti\'an, Spain}
\affiliation{Multiverse Computing, Paseo de Miram\'on 170, E-20014 San Sebasti\'an, Spain}

\author{Sukhbinder Singh}
\affiliation{Multiverse Computing, Spadina Ave., Toronto, ON M5T 2C2, Canada}

\author{Rom\'an Or\'us}
\affiliation{Donostia International Physics Center, Paseo Manuel de Lardizabal 4, E-20018 San Sebasti\'an, Spain}
\affiliation{Multiverse Computing, Paseo de Miram\'on 170, E-20014 San Sebasti\'an, Spain}
\affiliation{Ikerbasque Foundation for Science, Maria Diaz de Haro 3, E-48013 Bilbao, Spain}

\begin{abstract}
We show how quantum-inspired 2d tensor networks can be used to efficiently and accurately simulate the largest quantum processors from IBM, namely Eagle (127 qubits), Osprey (433 qubits) and Condor (1121 qubits). We simulate the dynamics of a complex quantum many-body system---specifically, the kicked Ising experiment considered recently by IBM in Nature 618, p. 500–505 (2023)---using graph-based Projected Entangled Pair States (gPEPS), which was proposed by some of us in PRB 99, 195105 (2019). Our results show that simple tensor updates are already sufficient to achieve very large unprecedented accuracy with remarkably low computational resources for this model. Apart from simulating the original experiment for 127 qubits, we also extend our results to 433 and 1121 qubits, \rom{and for evolution times around 8 times longer,} thus setting a benchmark for the newest IBM quantum machines. We also report accurate simulations for infinitely-many qubits. Our results show that gPEPS are a natural tool to efficiently simulate quantum computers with an underlying lattice-based qubit connectivity, such as all quantum processors based on superconducting qubits.      

\end{abstract}

\maketitle

\section{Introduction} We are currently witnessing an unprecedented technology race to develop practical large-scale quantum computers. While several hardware architectures have been developed, the largest available  quantum processors are those built with superconducting qubit technology \cite{scq}. In this setting, IBM's quantum roadmap is particularly promising, with the delivery of increasingly-larger quantum processors every year: Eagle with 127 qubits in 2021, Osprey with 433 qubits in 2022, and Condor with 1121 qubits and expected by the end of 2023. These are presently among the most powerful quantum machines worldwide. Furthermore, a large  effort is being devoted to mitigate errors in the processors, to become able to run longer quantum circuits and therefore increase quantum volume. Such error mitigation was pushed to an unprecedented level in a recent paper \cite{IBMNat}, where the IBM team simulated the dynamics of a kicked quantum Ising model on a 127-qubit 2d lattice that matched the connectivity topology of Eagle's quantum computer. These results are a great step forward  towards practical quantum computation in superconducting quantum processors. However, and unlike originally thought, they are still far from any sort of quantum advantage: as pointed out by several authors, the experiment can be simulated efficiently by purely classical means \cite{Miles, Sim2, Sim3, Sim4, Sim5, Sim6,Sim7}, and specially by methods using quantum-inspired tensor networks \cite{tn1, tn2}. 

In this paper we go one step further, and show how 2d tensor networks based on Projected Entangled Pair States (PEPS) \cite{PEPS, iPEPS1, iPEPS2} can be used to simulate IBM's largest quantum processors: Eagle  (127 qubits), Osprey (433 qubits) and Condor (1121 qubits). We show this by simulating the kicked Ising experiment mentioned above, with unprecedented accuracy and not just for 127 qubits, as in the original proposal, but for the larger quantum processors \rom{and longer evolution times}, setting new benchmarks for those machines. We use graph-based Projected Entangled Pair States (gPEPS) \cite{gPEPS}, a type of 2d tensor network algorithm which provides great flexibility in adapting to new lattices, both of finite and infinite size. We conclude that gPEPS is a natural tool to efficiently and accurately simulate slightly-entangled quantum computations on quantum computers that have an underlying lattice-based qubit connectivity.       

\section{Model} We implement a simulation of IBM kicked quantum Ising model. Specifically, we consider the dynamics generated by the spin-$1/2$ Hamiltonian 
\begin{eqnarray}
H = - J \sum_{\langle i,j \rangle} Z_i Z_j + h \sum_i X_i,    
\end{eqnarray}
with $Z_i, X_i$ being the $Z$ and $X$ Pauli matrices at site $i$, coupling $J$, transverse magnetic field $h$, and where the the sum of interactions is over nearest neighbors $\langle i, j \rangle$ on a lattice matching the topology of IBM's quantum processors. A first-order trotterization of the time evolution leads to the unitary operator 
\begin{eqnarray}
U(\theta_h) = \left( \prod_{\langle i,j \rangle} e^{i\frac{\pi}{4} Z_i Z_j}\right)\left( \prod_{i} e^{-i\frac{\theta_h}{2} X_i} \right),    
\end{eqnarray}
with $\theta_h$ a parameter controlling the relative strength of the field with respect to the spin-spin interaction. Starting from an initial state with all spins in the $\ket{0}$ state (i.e., all ``up"), we simulate the dynamics by applying the unitary operator $U(\theta_h)$ multiple times, therefore generating the state 
\begin{eqnarray}
|\psi(\theta_h,n)\rangle \equiv \left(U(\theta_h)\right)^n \ket{0}^{\otimes m},   
\end{eqnarray}
after $n$ applications of the operator, $m$ being the number of spins in the lattice.  

\section{Method}
\begin{figure}
    \includegraphics[width=0.485\textwidth]{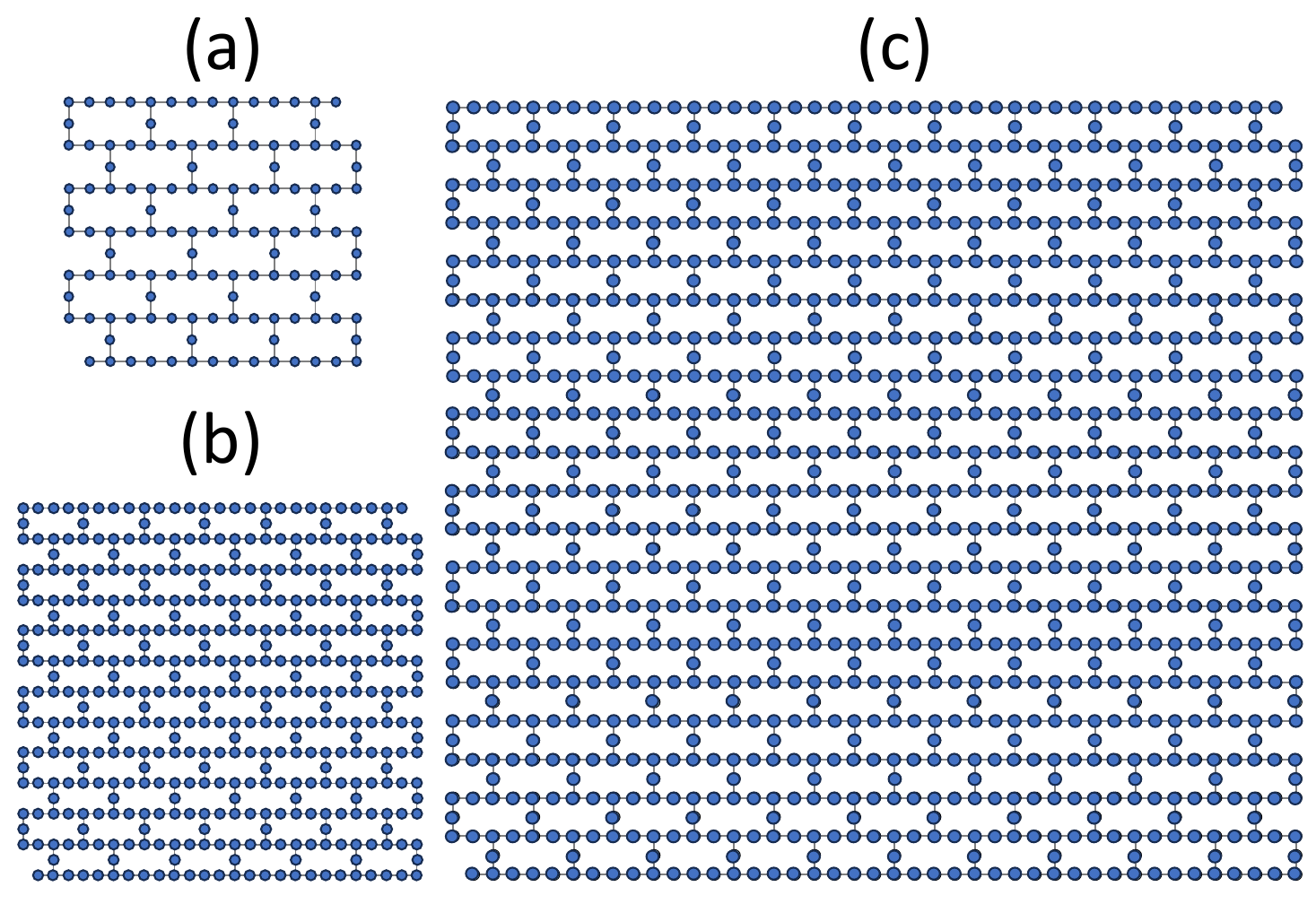}
    \caption{\justifying (Color online) Different heavy-hexagon lattices, corresponding to the topology of qubit connectivity of three IBM quantum processors: {\bf (a)} Eagle, with 127 qubits; {\bf (b)} Osprey, with 433 qubits; {\bf (c)} Condor, with 1121 qubits. Every dot in the lattices corresponds to a superconducting qubit, and every link corresponds to a qubit-qubit coupling.}
    \label{figlat}
\end{figure}

Here we simulate the dynamics of the above model on finite heavy-hexagon lattices with open boundary conditions and $127$, $433$ and $1121$ vertices, respectively matching the connectivity of qubits in Eagle, Osprey and Condor, see Fig.\ref{figlat}. For this, we adapt the gPEPS method \cite{gPEPS}---initially proposed for infinite systems---to finite-size lattices. We also use gPEPS to study the heavy-hexagon lattice in the thermodynamic limit with a unit cell of $10$ sites. 
The gPEPS algorithm is a quantum-inspired tensor network method that allows to easily simulate systems on generic lattices with desired dimensionality. As such, it is a natural evolution of the original iPEPS algorithm \cite{iPEPS1, iPEPS2} to simulate two-dimensional quantum lattice systems and also proposed years ago by some of us. gPEPS makes use of the \emph{simple update} of tensors \cite{simpleupdate, lubi, lubi2} and a \emph{mean-field approximation} for expectation values. These approximations are accurate for slightly-entangled 2d quantum lattice systems. Though they can be systematically improved, e.g., by using full and fast-full \cite{ff} updates and corner transfer matrices \cite{iPEPS2}, we have observed that the simplest of our approximations is already capable of simulating the system at hand with large accuracy. In our approach, the bond dimension of the PEPS tensor network is $\chi$, and is also the truncation parameter in our simulations: larger the $\chi$, the larger the allowed entanglement per bond. For comparison, we have also studied the effect of re-gauging the PEPS using belief propagation (BP) after each trotter step, as proposed in Ref.\cite{Miles}. 

\begin{figure*}[t]
\centering
\begin{minipage}{0.323\textwidth}
\centering
\begin{subfigure}{\linewidth}
\includegraphics[width=\linewidth]{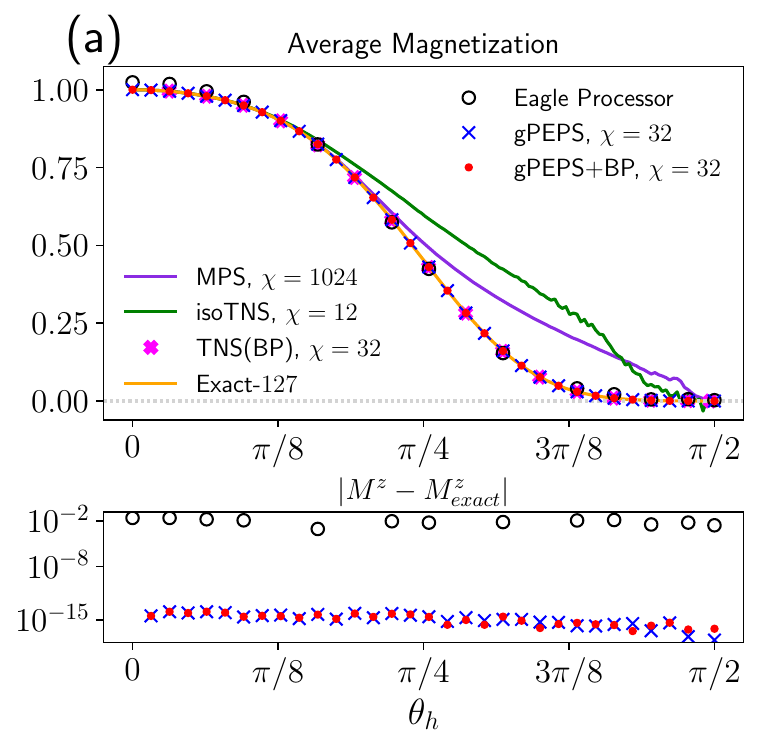}
\end{subfigure}
\end{minipage}
\hspace{0.\linewidth}
\begin{minipage}{0.323\textwidth}
\centering
\begin{subfigure}{\linewidth}
\includegraphics[width=\linewidth]{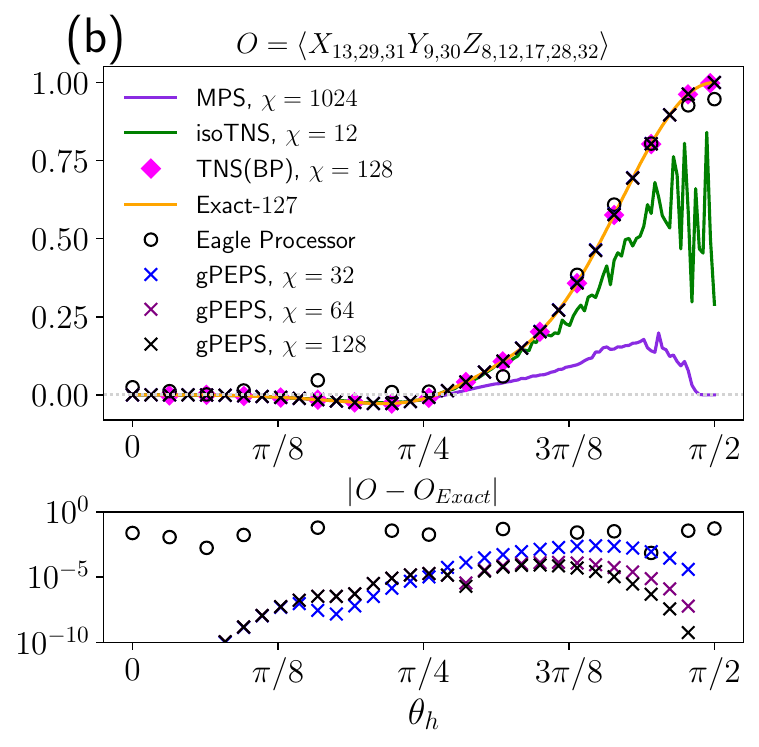}
\end{subfigure}
\end{minipage}
\hspace{0.0\linewidth} 
\begin{minipage}{0.323\textwidth}
\centering
\begin{subfigure}{\linewidth}
\includegraphics[width=\linewidth]{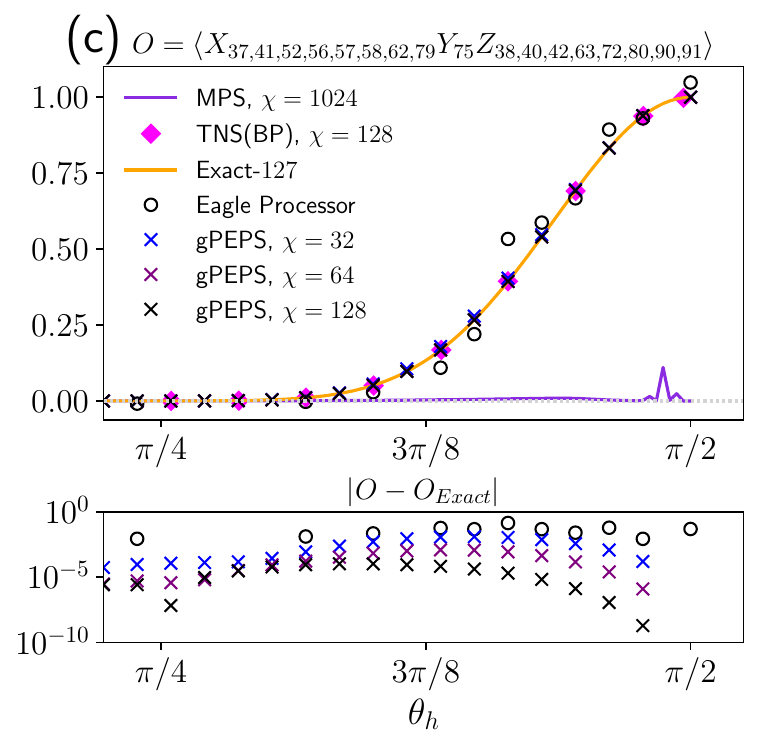}
\end{subfigure}
\end{minipage}
\vspace{-10pt}
\caption{\justifying (Color online) Comparing the gPEPS method in simulating the kicked transverse field Ising model against the 127-qubit IBM Eagle quantum processor and various other tensor network methods. The operator expectation values shown in \textbf{(a)} Average Magnetization, \textbf{(b)} Weight-10 observable, and \textbf{(c)} Weight-17 observable, are computed with respect to the state $|\psi(\theta_h,5)\rangle$. Each bottom plot shows the absolute difference between the light-cone based exact results and the results obtained through simulations (gPEPS and Eagle processor). Labelling of qubits is done sequentially, from left to right and top to bottom, starting with 0.}
\label{fig:3}
\end{figure*}

\section{Results}
\subsection{Benchmarking}
First, we simulate the 127-site heavy-hexagon lattice from Fig.\ref{figlat}(a). Using gPEPS we perform the unitary evolution $\left(U(\theta_h)\right)^n$ up to $n=5$ trotter steps, followed by computing expectation values  using a mean-field approximation. We reproduce Fig.3 from Ref.\cite{IBMNat} in our Fig.\ref{fig:3}, where we compare the outcome of our simulations with those obtained from experimental calculations performed on the IBM Eagle quantum processor. Additionally, we benchmark our findings against other tensor network methods. Comparison of our average magnetization values with the available light cone-based exact solution \cite{IBMNat} shows exceptional precision ($\sim 10^{-15}$ of absolute error), with each data point taking on average 2 seconds to run on a standard desktop PC {(Windows 11, Intel i7-11700 @2.50GHz, 16 GB RAM)}. Our results not only surpass the outcomes of IBM's quantum simulations, but they also outperform some of the best state-of-the-art tensor networks methods in both precision and speed. 

Additionally, to study the effect of Belief Propagation (BP) gauging we have independently simulated the unitary evolution of 5 trotter steps, where we do BP gauging after each trotter step. We find that BP does not improve accuracy \cite{SU_equal_BP}, as can be seen in Fig.\ref{fig:3}(a), even though the average computational time per point increased to 9.2 seconds. 

We also computed the expectation value of ``higher-weight'' observables  {(Appendix.\ref{sec:weight-N})} reported in the IBM experiment, as shown in Figs.\ref{fig:3}(b,c). Here, we have also included the tensor network results from Ref.\cite{Miles} for comparison. In these plots we provide the expectation values for the Weight-10 and Weight-17 operators, acting respectively on 10 and 17 lattice sites, across a range of $\theta_h$ values. The plots show that we obtain better precision than the quantum processor with a small bond dimension $\chi=32$, requiring an average compute time of 10 seconds per computed data point. As expected, we see an increase in accuracy by ramping up the bond dimension to $\chi=64$ and $\chi=128$ in Fig.\ref{fig:3}(c). 
 
However, in the range $\theta_h \in (\pi/8,\pi/4)$ in Fig.\ref{fig:3}(b), the Weight-10 observable expectation value with $\chi=32$ is found to be more accurate than the higher bond dimension results, the explanation can be found in the Appendix.\ref{sec:critical}.
\color{black}

\begin{figure*}[t] 
\centering
\begin{minipage}{0.323\textwidth}
\centering
\begin{subfigure}{\linewidth}
\includegraphics[width=\linewidth]{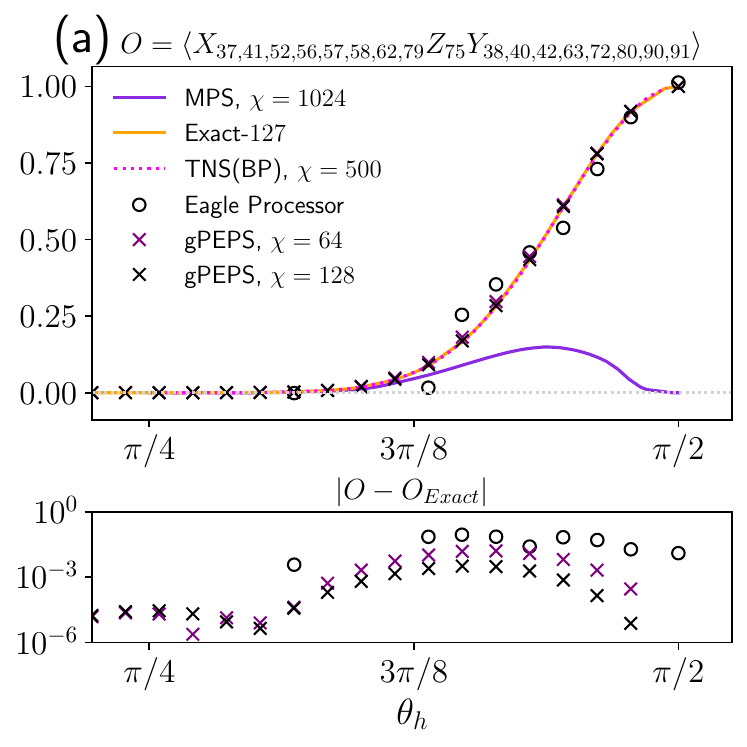}
\end{subfigure}
\end{minipage}%
\hspace{0.025\linewidth} 
\begin{minipage}{0.323\textwidth}
\centering
\begin{subfigure}{\linewidth}
\includegraphics[width=\linewidth]{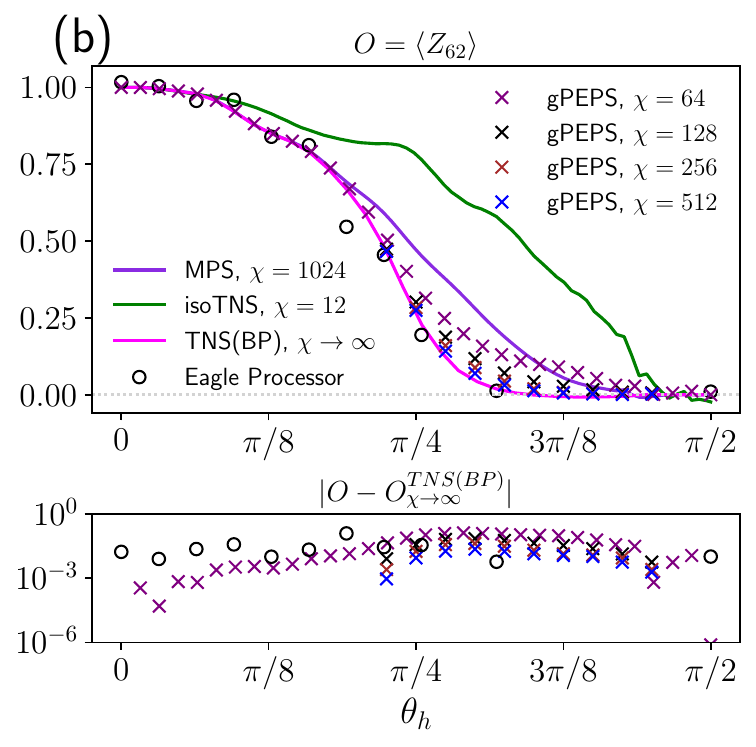}
\end{subfigure}
\end{minipage}%
\hspace{0.025\linewidth} 
\begin{minipage}{0.284\textwidth}
\centering
\begin{subfigure}{\linewidth}
\includegraphics[width=\linewidth]{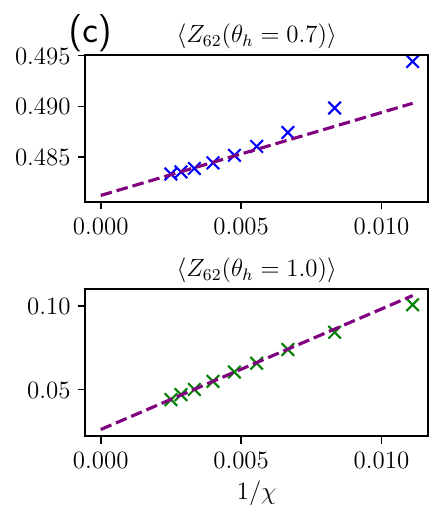}
\end{subfigure}
\end{minipage}
\caption{\justifying (Color online) Comparison of the gPEPS simulation with higher number of trotter steps with the Eagle quantum processor and various other tensor network methods. \textbf{(a)} Weight-17 observable computed after 6 trotter steps with respect to the state $|\psi(\theta_h,6)\rangle$. The bottom plot shows the absolute difference between our simulation and the available exact result. \textbf{(b)} Weight-1 expectation value computed after 20 trotter steps with respect to the state $|\psi(\theta_h,20)\rangle$. Because of the absence of exact result for this simulation, we have computed the absolute difference between our simulation and the BP-approximation tensor network state approach with $\chi\rightarrow \infty$, presented in the bottom subplot. \textbf{(c)} Finite-entanglement scaling of Weight-1 observable expectation value $\langle \psi(\theta_h,20)|  Z_{62} |\psi(\theta_h,20) \rangle$ with respect to the inverse of bond dimension ($1/\chi$) for two distinct $\theta_h$ values. Labeling of qubits is done sequentially, from left to right and top to bottom, starting with 0.}
\label{fig:4}
\end{figure*}

Next, we have studied the case in which the unitary evolution spans more than 5 trotter steps. This involves simulating the state corresponding to the extended-depth quantum circuit, as shown in Figure 4 of Ref.\cite{IBMNat}. We computed the Weight-17 observable after 6 trotter steps and compare it with the result obtained by the 127-qubit quantum processor, and our results can be found on Fig.\ref{fig:4}(a). Comparison against the exact result \cite{IBMNat} shows that gPEPS simulations with bond dimension $\chi=64$ already outperform the Eagle quantum processor in accuracy. As expected, we have observed further enhancements in accuracy for larger bond dimension $\chi = 128$. 

\begin{figure*}[t] 
\centering
\begin{minipage}{0.323\textwidth}
\centering
\begin{subfigure}{\linewidth}
\includegraphics[width=\linewidth]{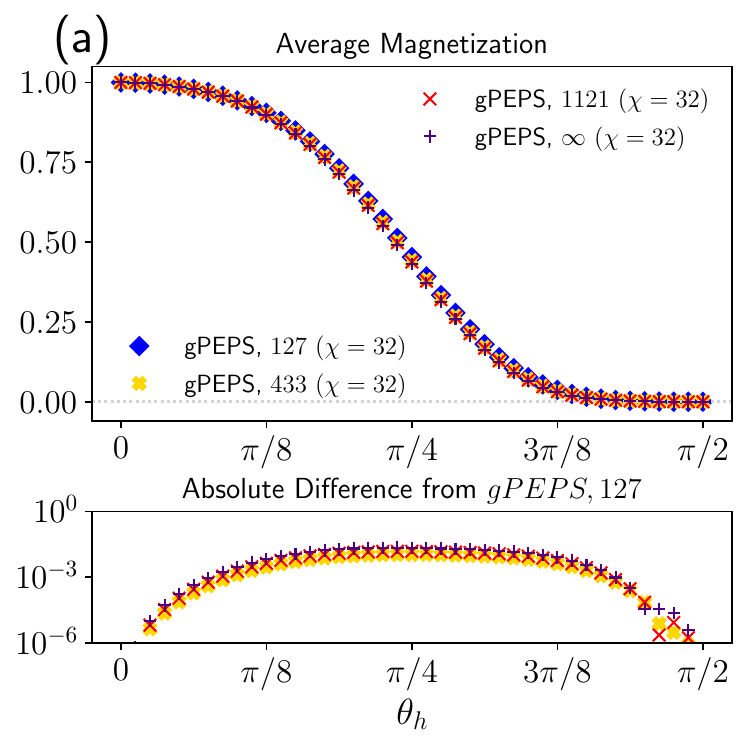}
\end{subfigure}
\end{minipage}%
\begin{minipage}{0.323\textwidth}
\centering
\begin{subfigure}{\linewidth}
\includegraphics[width=\linewidth]{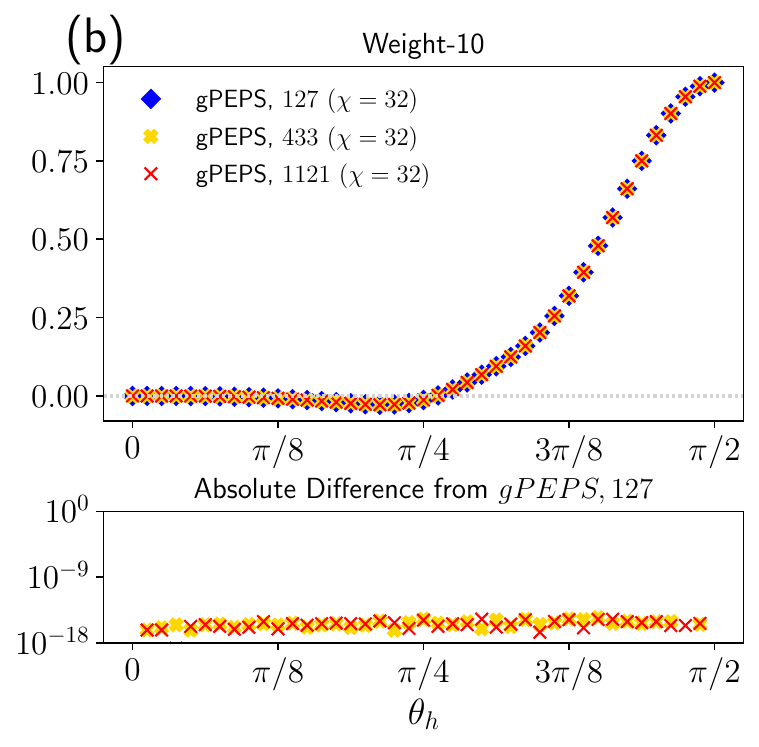}
\end{subfigure}
\end{minipage}%
\begin{minipage}{0.323\textwidth}
\centering
\begin{subfigure}{\linewidth}
\includegraphics[width=\linewidth]{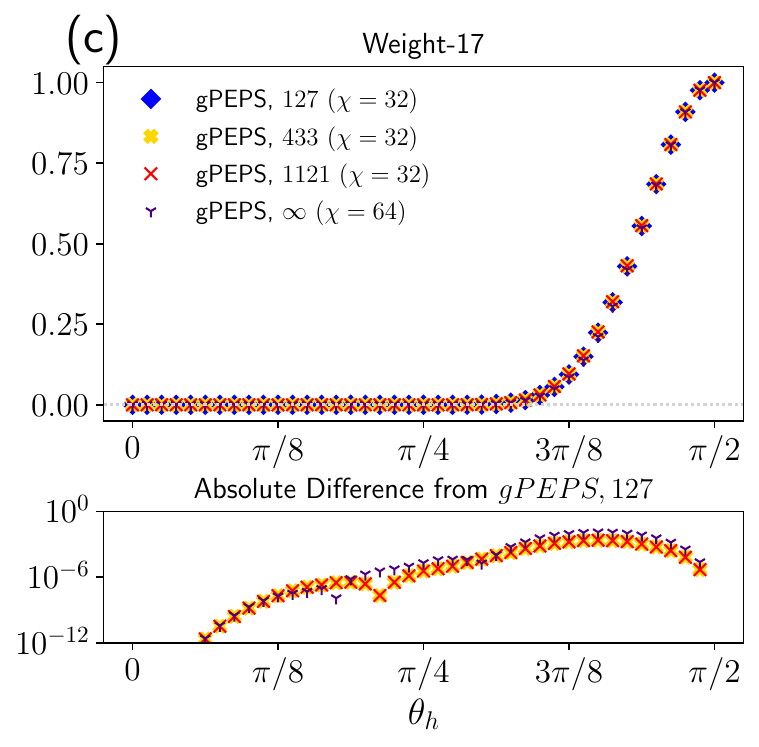}
\end{subfigure}
\end{minipage}
\caption{\justifying (Color online) Results of simulating various IBM quantum chips with higher number of qubits using gPEPS: Eagle processor with 127 qubits, Osprey with 433 qubits, Condor with 1121 qubits, and the heavy-hexagon lattice in thermodynamic limit. \textbf{(a)} Average magnetization, \textbf{(b)} Weight-10 observable near the open edge, \textbf{(c)} Weight-17 observable deep inside the bulk. The structure of the Weight-10 and Weight-17 observables is discussed in \textcolor{blue}{(Appendix.\ref{sec:weight-N})}.}
\label{fig:Fig3_sizes}
\end{figure*}

To further test the algorithm we performed simulations involving longer unitary evolutions, with 20 trotter steps, and across a range of $\theta_h$ values. We computed the expectation value of the Weight-1 (single-site) operator, as shown in Fig.\ref{fig:4}(b). Notably, we achieved numerically exact results for the Clifford points $\theta_h=0$ and $\theta_h=\pi/2$ when using bond dimension $\chi=64$. While an exact solution for longer unitary evolution remains elusive, we were able to compare with tensor network results in the infinite bond dimension limit \cite{Miles}, obtained from finite-entanglement scaling. While our $\chi=64$ bond dimension accurately captures points for $\theta_h\lesssim 3\pi/16$, noticeable deviations become apparent beyond this regime. As shown in the figure, we could improve the accuracy significantly by increasing the bond dimension to $\chi=128, 256$ and $512$. Finally, to show the reliability of our method we have studied the finite-entanglement scaling of  $\langle Z_{62} \rangle (\theta_h)$ with the inverse bond dimension ($1/\chi$), as shown in Fig.\ref{fig:4}(c) for $\theta_h=1.0$ and $\theta_h=0.7$. We have plotted the values for 9 different bond dimensions and the fitting  of extrapolation is done for the 5 highest bond dimensions. While for $\theta_h=0.7$ we start to see a tendency to saturation at large bond dimension, at $\theta_h=1.0$ we see no clear evidence yet of bond dimension saturation, which indicates that the quantum state has large entanglement. 

\subsection{Large systems} 
Next, and thanks to the computational efficiency of gPEPS, we have successfully simulated larger IBM quantum systems involving 433 (Osprey) and 1121 (Condor) qubits, corresponding to the heavy-hexagon lattices in Figs.\ref{figlat}(b,c). While the original IBM experiment in Ref.\cite{IBMNat} was implemented only for 127 qubit system, we understand our results as a benchmark for future experiments on these larger quantum processors. In addition, we have also implemented gPEPS for the system with infinitely-many qubits, by assuming translation invariance in the PEPS tensor network with a unit cell of 10 sites. For all these sizes, we have simulated the unitary evolution of 5 trotter steps and computed a number of observables. Our results can be found in Fig.\ref{fig:Fig3_sizes}. First, we computed the average magnetization in the $Z$-direction for $\chi = 32$, as shown in Fig.\ref{fig:Fig3_sizes}(a). We can see that there are minimal differences for all sizes (127, 433, 1121 and infinite), indicating that the bulk of the system is already quite close to the thermodynamic limit already for 127 qubits. For this plot, the average simulation time of one data point for sizes 127, 433, 1121 and infinite are respectively 2, 8.3, 50, and 0.17 seconds. Next, in order to test possible boundary effects, we compare in Fig.\ref{fig:Fig3_sizes}(b) the results for a Weight-10 observable near the open boundary of the heavy-hexagon lattice, for the three finite-size systems, again for $\chi = 32$. The composition of the observable and its calculation is discussed in the \textcolor{blue}{Appendix.\ref{sec:weight-N}}. As we can observe, there is no appreciable difference in the result, signalling again that even the smallest lattice is already close to the thermodynamic limit. Last but not least, we have also computed the expectation value of a Weight-17 operator deep inside the bulk of the system, for all lattices (including the infinite one for $\chi = 64$), and the results are in Fig.\ref{fig:Fig3_sizes}(c). Minimal difference among the results of different system sizes show that 127 qubit system is already very close to the thermodynamic limit.

\begin{figure*}[t]
\centering
\begin{minipage}{0.323\textwidth}
\centering
\begin{subfigure}{\linewidth}
\includegraphics[width=\linewidth]{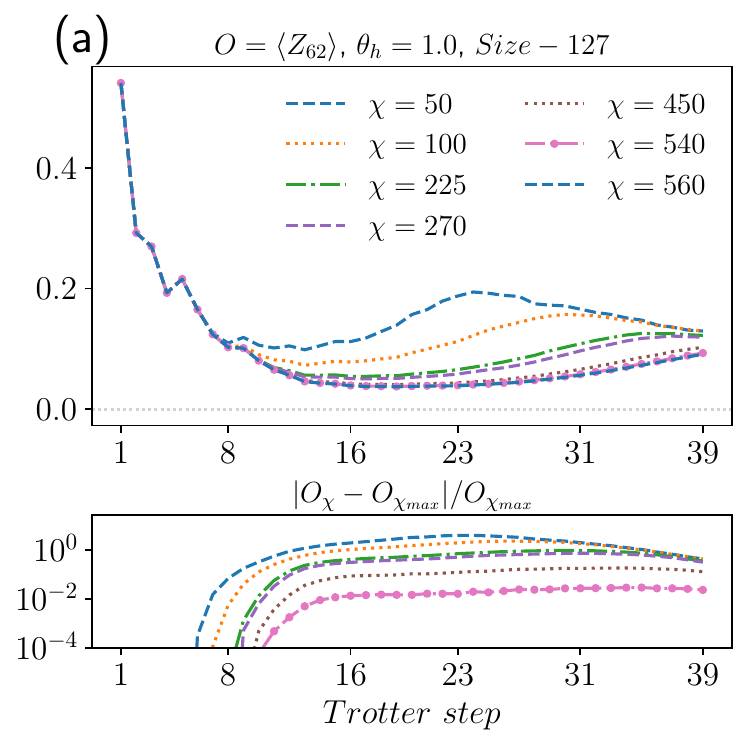}
\end{subfigure}
\end{minipage}
\hspace{0.\linewidth}
\begin{minipage}{0.323\textwidth}
\centering
\begin{subfigure}{\linewidth}
\includegraphics[width=\linewidth]{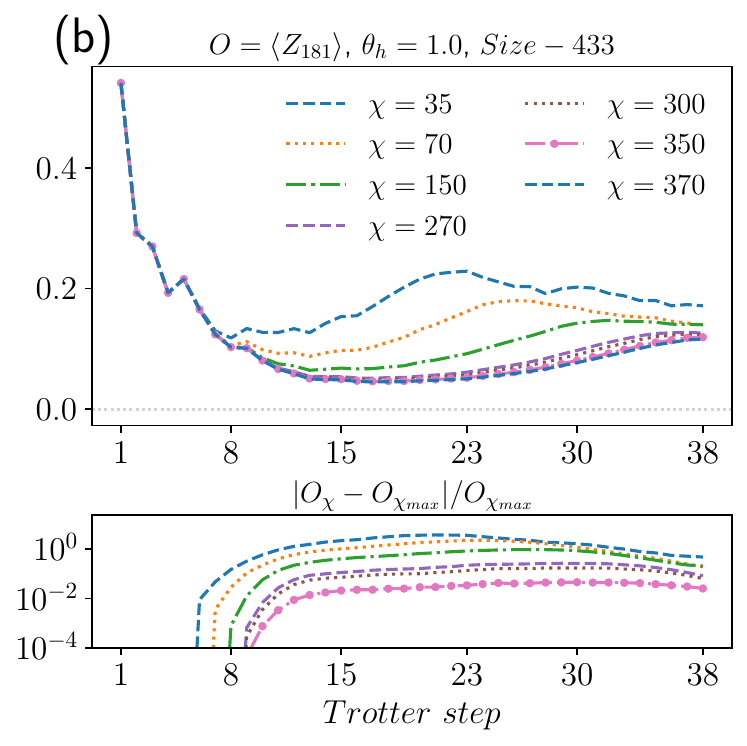}
\end{subfigure}
\end{minipage}
\hspace{0.0\linewidth} 
\begin{minipage}{0.323\textwidth}
\centering
\begin{subfigure}{\linewidth}
\includegraphics[width=\linewidth]{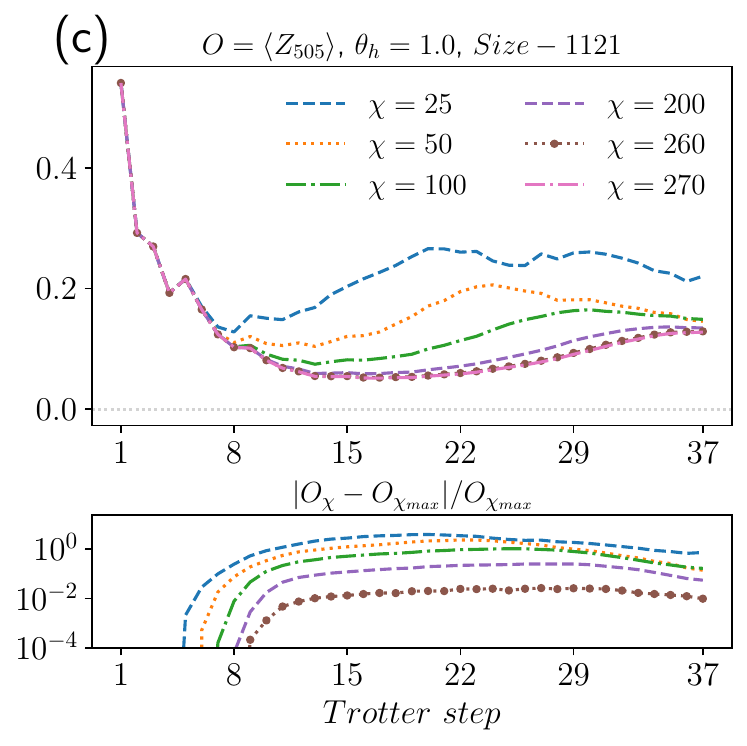}
\end{subfigure}
\end{minipage}
\vspace{-10pt}
\caption{\justifying (Color online) Long time evolution of the magnetization for a site in the bulk at $\theta_h = 1.0$ and the three different sizes: {\bf (a)} 127 qubits, up to $\chi = 560$ and 39 Trotter steps, {\bf (b)} 433 qubits, up to $\chi = 370$ and 38 Trotter steps, {\bf (c)} 1121 qubits, up to $\chi = 270$ and 37 Trotter steps. Lower panel shows relative errors with respect to the maximum achievable bond dimension.}
\label{fig:5}
\end{figure*}

\begin{figure*}[t]
\centering
\begin{minipage}{0.323\textwidth}
\centering
\begin{subfigure}{\linewidth}
\includegraphics[width=\linewidth]{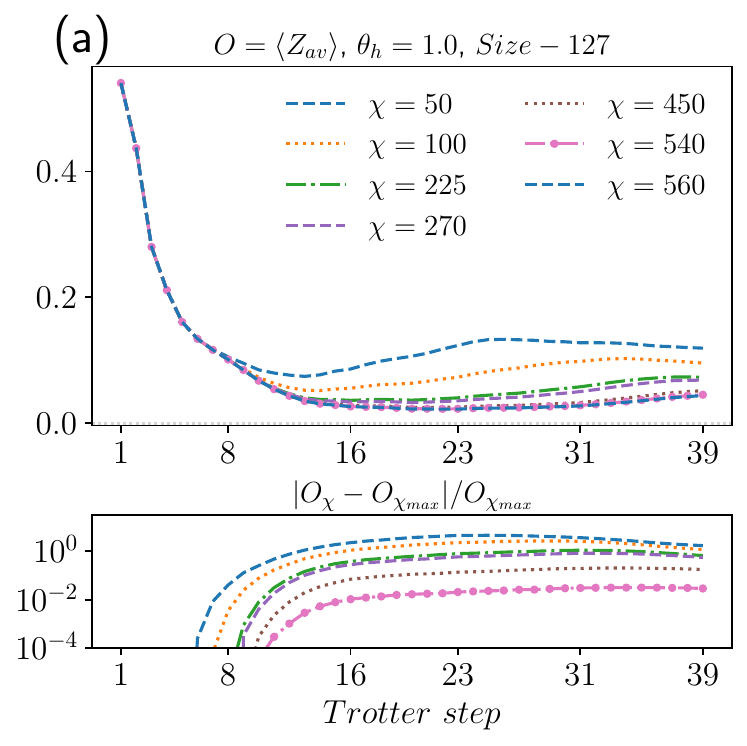}
\end{subfigure}
\end{minipage}
\hspace{0.\linewidth}
\begin{minipage}{0.323\textwidth}
\centering
\begin{subfigure}{\linewidth}
\includegraphics[width=\linewidth]{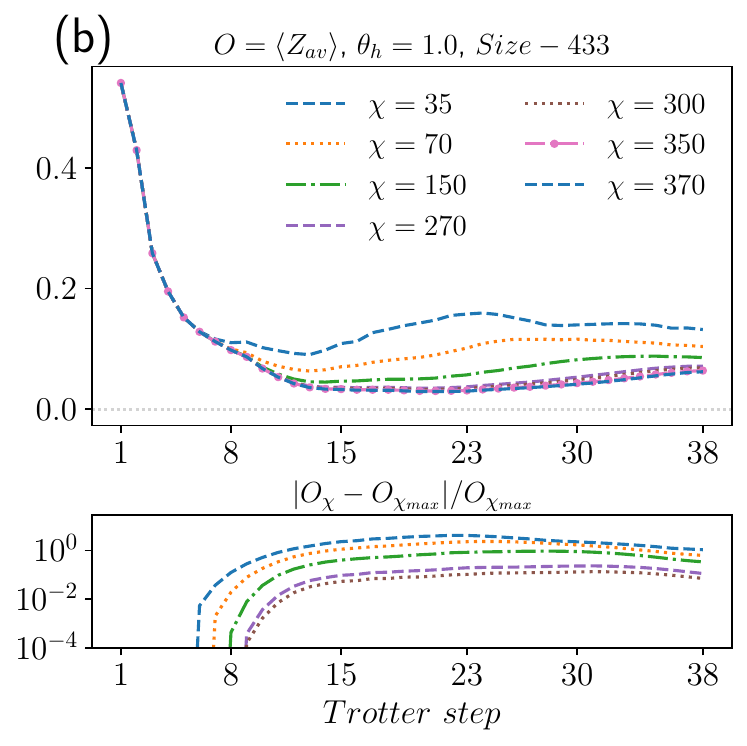}
\end{subfigure}
\end{minipage}
\hspace{0.0\linewidth} 
\begin{minipage}{0.323\textwidth}
\centering
\begin{subfigure}{\linewidth}
\includegraphics[width=\linewidth]{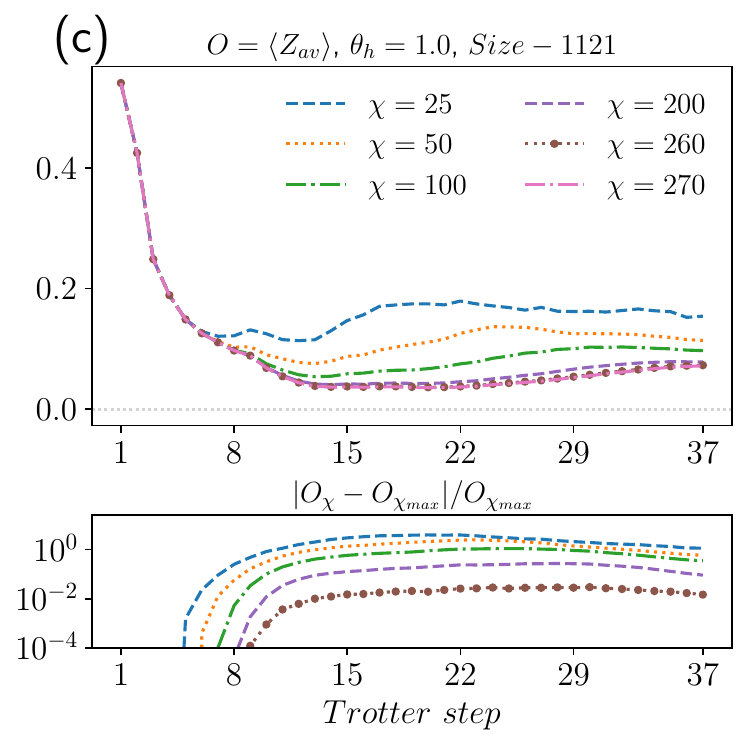}
\end{subfigure}
\end{minipage}
\vspace{-10pt}
\caption{\justifying (Color online) Long time evolution of the average magnetization of the whole system at $\theta_h = 1.0$ and the three different sizes: {\bf (a)} 127 qubits, up to $\chi = 560$ and 39 Trotter steps, {\bf (b)} 433 qubits, up to $\chi = 370$ and 38 Trotter steps, {\bf (c)} 1121 qubits, up to $\chi = 270$ and 37 Trotter steps. The lower panel shows relative errors with respect to the maximum achievable bond dimension.}
\label{fig:6}
\end{figure*}

\begin{figure}[!h]
\includegraphics[width = 0.4\textwidth]{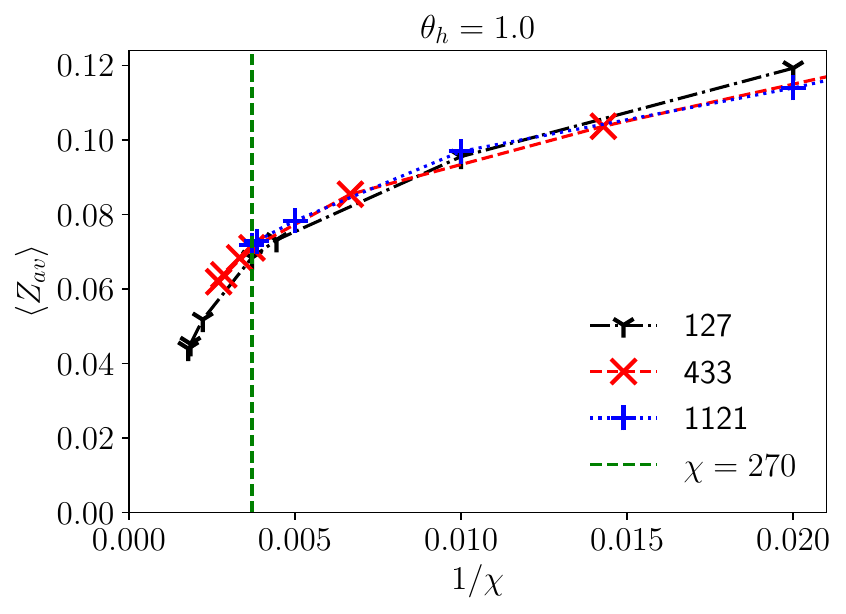}
\caption{\justifying {(Color online) Bond dimension scaling of the average magnetization for different system sizes.}}
\label{fig:ChiScaling}
\end{figure}

\subsection{Longe time evolution}
\rom{All the above results motivate us to test the limit of our simulation method. One should expect that long-time evolutions may create a large amount of entanglement that is hard to be captured by the gPEPS technique. This could be captured as a loss of convergence with the bond dimension $\chi$ in our simulations, setting then a benchmark: a quantum computer claiming quantum advantage in simulating this model should (at the very least) be able to compute time evolutions longer than those for which gPEPS loses convergence. Therefore, to understand the limit of our method we computed the results in Figs.(\ref{fig:5}) and (\ref{fig:6}), respectively for the time evolution of the magnetization of a site deep in the bulk and the average magnetization over all sites, for the three considered sizes (127, 433 and 1121 qubits). The results are for a large number of Trotter steps (between 37 and 39), and for the largest bond dimension that we could achieve for each size.  
$\chi=270$ is the maximum common bond dimension that we could simulate for all three system sizes. Bond dimension scaling is shown in Figs.\ref{fig:ChiScaling}.
\color{black}
As we can see in the plots, for all sizes we find the same result: the study of consecutive bond dimensions shows that \emph{the observables converge even for a large number of Trotter steps}. This is indeed surprising and seems to indicate that the gPEPS technique is particularly suited to capture the entanglement structure of the heavy hexagon lattice. The reason behind this may be that this lattice, after all, can be quite well approximated by a tree-like structure with no loops {(Appendix.\ref{sec:localtree})}. And the properties of such loop-free structures can be captured with large precision by the simple tensor update that we use in gPEPS.} 

\section{Conclusions} 
In this paper we have simulated IBM's kicked Ising experiment \cite{IBMNat} on heavy-hexagon lattices with 127, 433, 1121, and infinitely-many qubits, using a quantum-inspired tensor network technique tailored to higher-dimensional systems. Our method uses the so-called gPEPS algorithm, which is remarkably efficient and accurate. Our method not only reproduces the results of the original experiment for 127 qubits but also settles new benchmarks for \rom{large quantum circuits on} IBM's Eagle, Osprey, and Condor quantum processors. We conclude that gPEPS is a natural tool to efficiently and accurately simulate slightly entangled quantum computations on quantum computers with an underlying lattice-based qubit connectivity, be it in 2 or higher dimensions, going much beyond the capabilities of other tensor network structures. In particular, it is an ideal tool to classically simulate quantum computers based on superconducting qubits. A relevant question triggered by our results is whether quantum processors based on artificial qubits (e.g., superconducting qubits, quantum dots, etc.), and with an underlying lattice-based connectivity, can reach a sufficiently-low noise level so that they cannot be  simulated classically by some tailored tensor network algorithm. It would also be interesting to assess gPEPS in the simulation of other types of quantum hardware with all-to-all qubit connectivity, such as trapped ions and neutral atoms.  
 
\section{Acknowledgments}
We acknowledge DIPC, Ikerbasque, Basque Government, Diputaci\'on de Gipuzkoa and EIC for support, as well as discussions with the teams from Multiverse Computing and DIPC. We also acknowledge Miles Stoudenmire and Joseph Tindall for providing access to previous data, and also comments by the community. S. S. J. also acknowledges IASBS. \rom{Simulations in Figs.(\ref{fig:4}(b),\ref{fig:5},\ref{fig:6}) were done in the ATLAS cluster.}


\appendix

\section{Results near critical point}
\label{sec:critical}
In the range of $\theta_h \in (\pi/8,\pi/4)$ depicted in Fig 2(b), our gPEPS results with lower bond dimension are more accurate than the higher ones, contrary to the monotonic improvement of accuracy with the increament of bond dimension. The underlying reason is the presence of a critical point ($\theta_h \approx 0.6$) in the vicinity, where the correlation length diverges. Our simple update-based gPEPS approach, along with local measurements, struggles to capture the correlations adequately with smaller bond dimensions, leading to overestimation or underestimation. Nonetheless, we have confirmed the convergence of our observable expectation values towards higher bond dimension.
\color{black}

\section{Weight-$N$ observables}
\label{sec:weight-N}

In the process of measuring the onsite observable, we use the mean-field approximation of the environment within the gPEPS algorithm. We provide comprehensive details of this approach in the main paper. However, measuring an observable that involves multiple particles with intricate loop structures poses a considerable challenge when aiming for optimal measurements. To address this challenge, we leverage the special Clifford property of the circuit at $\theta_h=\pi/2$. This strategic choice allows us to transform the problem of computing higher-weight observables into a more tractable task: measuring a  Weight-1 observable but with a higher number of trotter steps involved. Though the computational cost increases with the number of trotter steps, in this way, we can use local measurements only to get an accurate value of the higher-weight observables. For example, in the case of the 127-size system, the Weight-10 and Weight-17 operators of Fig.\ref{fig:3} are given by 
\beqa
W^{(127)}_{10}&=&X_{13,29,31} Y_{9,30} Z_{8,12,17,28,32}, \nonumber \\ 
W^{(127)}_{17}&=&X_{37,41,52,56,57,58,62,79} Y_{75} Z_{38,40,42,63,72,80,90,91}. \nonumber\\
\eeqa
The expectation value of these operators with respect to the state $|\psi(\theta_h,5)\rangle$ (obtained after $5$ trotter steps) can be re-written  at the Clifford point as 
\begin{eqnarray}
\langle  W^{(127)}_{10} \rangle_5 &=& \langle \psi(\theta_h,5) | U^5(\pi/2) Z_{13} (U^{\dagger}(\pi/2))^5 | \psi(\theta_h,5) \rangle \nonumber \\
\langle  W^{(127)}_{17} \rangle_5 &=& \langle \psi(\theta_h,5) | U^5(\pi/2) Z_{58} (U^{\dagger}(\pi/2))^5 | \psi(\theta_h,5) \rangle. \nonumber\\
\end{eqnarray}
As a result, to determine e.g. the Weight-10 expectation value after $n$ trotter steps, one may just compute the single site expectation value of $\langle Z_{13} \rangle$ with respect to the state 
\begin{eqnarray}
|\omega(\theta_h,n)\rangle &=&  \left( U^{\dagger}(\pi/2) \right)^n | \psi(\theta_h,n) \rangle.
\end{eqnarray}
To obtain the state $|\omega(\theta_h,n)\rangle$ we evolve the $|\psi(\theta_h,n)\rangle$ with the operator $U^{\dagger}(\pi/2)$ for $n$ trotter steps. We have used this approach to compute observables for all lattice sizes, namely 127, 433, 1121 and infinite. In particular, we compute the expectation values 
\begin{eqnarray}
\langle  W^{(size)}_{10} \rangle_5 &=&  \langle \omega(\theta_h,5) |Z_{P(size)} | \omega(\theta_h,5) \rangle, \nonumber\\
\langle  W^{(size)}_{17} \rangle_5 &=&  \langle \omega(\theta_h,5) |Z_{Q(size)} | \omega(\theta_h,5) \rangle.  
\end{eqnarray}
In the above equations, integers $P(size)$ and $Q(size)$ represent qubit labels depending on the size, as shown in the table below. Vertices are labeled sequentially from left to right and top to bottom, starting with 0.

\begin{table}[!h]
\centering
\begin{tabular}{|c|c|c|c|c|}
\cline{2-5}
\multicolumn{1}{c|}{size $\rightarrow$ } & \textbf{127} & \textbf{433} & \textbf{1121} & \textbf{Infinity} \\ \hline
\textbf{P} & 13 & 25 & 41 & -- \\ \hline
\textbf{Q} & 62 & 181 & 505 & 2 \\ \hline
\end{tabular}
\caption{Relation of qubit labels for different sizes, for the Weight-10 and Weight-17 observables of Fig.\ref{fig:Fig3_sizes}.}
\label{tab:PQ}
\end{table}

\section{Local tree structure of heavy-hexagon lattice}
\label{sec:localtree}

\begin{figure} [!h]
    \includegraphics[width=0.485\textwidth]{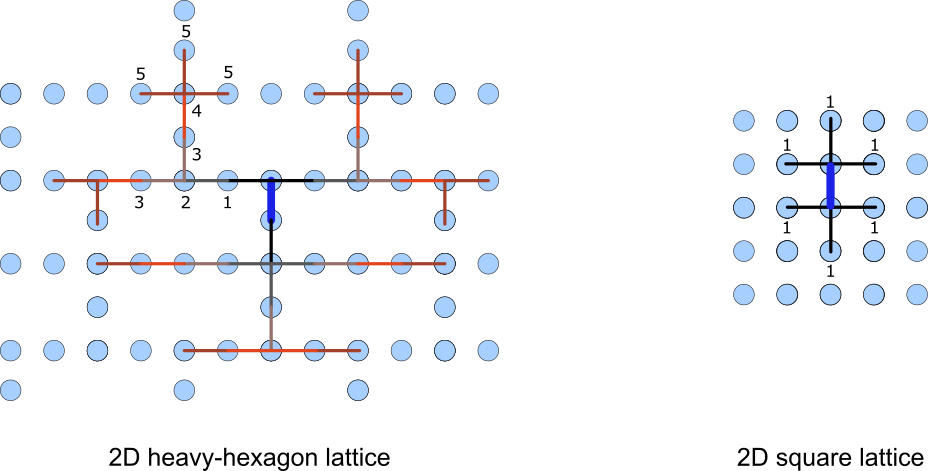}
    \caption{\justifying (Color online) Comparing a patch of heavy-hexagon lattice (left) with square lattice (right).}
    \label{figtree}
\end{figure}
Simple update and local measurements are most effective for lattices exhibiting a local tree structure and shorter-range correlations, owing to the absence of loops. In the Figs.\ref{figtree}, we present a comparison between a patch of 2D heavy-hexagon lattice and a 2D square lattice, illustrating the environment of the blue edge (pair of spins). It is evident that for a 2D square lattice, the `n-th'-neighbor environment contains loops for $n>1$. Conversely, in the case of a heavy-hexagonal lattice, the `n-th'-neighbor environment exhibits loops for $n>5$. Thus, for models with short-range correlations, such as the Ising transverse field away from the critical point, the heavy-hexagon lattice behaves akin to a tree structure locally. Consequently, local tensor updates employed in gPEPS, as well as local measurements, excel in capturing the real-time dynamics of the kicked Ising experiment.
\color{black}

\bibliography{bibliography}{}

\end{document}